\documentclass{article}
\begin{document}
\large

\centerline{\Large \bf The Stability of an Isentropic Model for a Gaseous Relativistic Star}

\vskip 0.9cm

\begin{center}
\centerline {\large P. S. Negi \footnote{e-mail: psnegi\_nainital@yahoo.com; negi@aries.ernet.in}} 

{\large Department of Physics, Kumaun University, Nainital-263 002, India}

\vspace {1.8cm}
{\large \bf Abstract}
\end{center}

\vspace {0.8cm}

\noindent   
We show that the isentropic subclass of Buchdahl's exact solution for a gaseous relativistic star is stable and gravitationally bound for all values of the compactness ratio $u [\equiv (M/R)$, where $M$ is the total mass and $R$ is the radius of the configuration in geometrized units] in the range, $0 < u \leq 0.20$, corresponding to the {\em regular} behaviour of the solution. This result is in agreement with the expectation and opposite to the earlier claim found in the literature.

\bigskip
keywords: {Static Spherical Structures: Pulsational Stability, Exact Gaseous Solutions}
\vspace {4.5cm}
\pagebreak

\noindent {\large \bf 1.\,\,\, Introduction}
\bigskip

 Inspite of non linear and coupled differential equations, various  exact  
solutions of Einstein's   field    equations for static and spherically symmetric perfect fluid sphere are  
available  in  the  literature [1]. But the problem of finding solutions corresponding to a {\em regular} (i.e. the finite positive density at the origin which decreases monotonically outwards) and {\em gaseous} (i.e. the density also vanishes at the surface of the configuration together with pressure) sphere is rather more difficult. Such solutions are, therefore, severely restricted to a limited number. To the knowledge of the present author, there exist only two exact solutions in this category which can serve as realistic models of relativistic stars. The first solution is Tolman's type VII solution with vanishing surface density [2, 3] and other one is Buchdahl's gaseous model [4, 5, 6]. Since such solutions may also find application to construct the models of relativistic supermassive stars [7], together with the usual models of neutron stars [8], their investigation is important. Even more important is the study of their gravitational binding and stability under small radial pulsations. 

The stability of Tolman's type VII solution with vanishing surface density has been undertaken in detail by Negi and Durgapal [7, 9] and they have shown that this solution remains gravitationally bound (for all permissible values of $u \leq 0.3861 $) and pulsationally stable (for all values of $u \leq 0.3428$). The pulsational stability of Buchdahl's gaseous solution was studied by Knutsen [6] and he found that this solution turns out to be unstable for the range of compactness ratio $0 \leq u \leq (1/6)$ (the range in which the Buchdahl's gaseous solution satisfies the regularity and causality conditions).
However, this result seems to be unlikely, because from the study of the behaviour of this solution, one finds that the solution becomes a classical polytrope of index 1 as $u \rightarrow 0$, whereas when $u = (1/6)$ the solution is equivalent to a sphere of homogeneous energy-density in the core (Schwarzschild core) surrounded by an envelope which becomes polytropic ($n = 1$) as the surface is reached. Since both of these solutions, the classical polytrope of index 1 and the homogeneous density sphere, represent the stable configurations in general relativity [10], it looks surprising that Buchdahl's gaseous model which becomes one of these solution in the limiting case, $u \rightarrow 0$, and the `effective' core-envelope model comprising both of these solutions as $u \rightarrow (1/6)$ is unstable for $0 \leq u \leq (1/6)$ ! 

In order to resolve the inconsistency between the reasoning presented above and the results of Knutsen [6], we also investigate the pulsational stability of Buchdahl's gaseous model on the basis of variational method [11,12] by using the general form of the `trial function', frequently used in the literature [10,13](whereas Knutsen [6] has used only a particular form of the `trial function' which also satisfies the boundary conditions). For comparison, we also employ the particular form of the `trial function' used by Knutsen [6]. Furthermore, we also consider the gravitational binding of the configuration which is necessary for the existence of a realistic structure (Knutsen [6] did not consider the gravitational binding of the structures). We perform our calculations for the range $0 \leq u \leq 0.20$ (we extend the range of our calculations for the interval $(1/6) < u \leq 0.20$ where the adiabatic sound speed, $v_s(=\sqrt (dP/dE))$, exceeds the speed of light, $c$. Actually, we are interested to investigate the stability for the full range in which the solution behaves like a regular solution and not just for the interval in which the `causality' is not violated. Since for the case of high densities, the adiabatic sound speed, $v_s$, does not necessarily represent the `true' signal propagation speed so that the `causality' may not be violated even if $v_s > c$ [14]). Our calculations show that perfect fluid subclass of Buchdahl's gaseous solution is stable and gravitationally bound for all values of $u$ in the range $0 \leq u \leq 0.20$.

\bigskip
\noindent {\large \bf 2.\,\,\, Metric and Expressions for Buchdahl's Gaseous Model}
\bigskip

The interior geometry of static spherically symmetric perfect fluid configurations in canonical coordinates is given by the metric 

\begin{equation}
ds^2 =  e^{\nu} dt^2 - e^{\lambda} dr^2 - r^2(d\theta^2 + r^2 {\rm sin}^2 \theta d\phi^2) 
\end{equation}

where $\nu$ and  $\lambda$ are functions of $r$ alone. 
Recalling that we are using the `geometrized units' [$G = c = 1$; where $G$ is the Universal gravitation constant and $c$ is the speed of light in vacuum], the metric coefficients of Eq.(1) for Buchdahl's gaseous solution yield in the following form [5,6] \footnote[1]{We use notation for parameters, different from those used in [5] and [6], for convenience}
\begin{equation}
e^{\nu} = (1 - 2u)\frac{\bar{c} - \bar{a}}{\bar{c} + \bar{a}}
\end{equation}

\begin{equation}
e^{\lambda} = 4(1 - 2u)\frac{\bar{c} + \bar{a}}{(\bar{c} - \bar{a})(\bar{c} + \bar{b})^2}
\end{equation}
The expressions for pressure, $P$, and energy-density, $E$, now read

\begin{equation}
8\pi PR^2 = \frac{\pi^2(1 - u)^2}{(1 - 2u)} \frac{\bar{a}^2}{(\bar{c} + \bar{a})^2}
\end{equation}

\begin{equation}             
8\pi ER^2 = \frac{\pi^2(1 - u)^2}{(1 - 2u)} \frac{\bar{a}(2\bar{c} - 3\bar{a})}{(\bar{c} + \bar{a})^2}
\end{equation}
 
where the variables $\bar{a}, \bar{b}$ and  $\bar{c}$ are defined as

\begin{equation} 
\bar{a}  = 2u({\rm sin}(z)/z)
\end{equation}

\begin{equation}
\bar{b}  = 2u\, {\rm cos}\, z
\end{equation}

\begin{equation}
\bar{c}  =  2(1 - u)
\end{equation}

and $z$ is given by
\begin{equation}
z = \frac {\pi y}{[1 + u({\rm sin}(z)/z)(1 - u)^{-1}]};\,\,\, 0 \leq z \leq \pi
\end{equation}
here $y$ is the dimensionless radial coordinate $(r/R)$ (i.e. the radial coordinate measured in units of configuration size) varies in the range $0 \leq y \leq 1$.

Eq.(9) yields
\begin{equation}
\pi dy = [1 + u\, {\rm cos}\, z(1 - u)^{-1}]dz
\end{equation}
The adiabatic sound speed, ${v_s}^2$, is given by
\begin{equation}
{v_s}^2 = \frac{dP}{dE} = \frac{\bar{a}}{\bar{c} - 4\bar{a}}
\end{equation}

Since the Eqs.(2)-(5) yield the pulsationally stable structures (section 4) for all permissible values of $u$, one can construct the models of physical objects (i.e., neutron stars and supermassive stars) on the basis of Buchdahl's gaseous solution. For an assigned value of the compactness ratio, $u$, and a suitable value of the central energy-density, $E_0$ (the value of $E$ at $r=0$), the radius $R$(and hence the total mass $M(=Ru)$) of the configuration can be obtained by using the following relation
\begin{equation}
R^2 = \frac{\pi^2u(2 - 5u)(1 - u)^2}{8\pi E_0(1 - 2u)}
\end{equation}

\bigskip

\bigskip  
\noindent {\large \bf 4.\,\,\,Dynamical Stability of Buchdahl's Gaseous Model}
\bigskip

\bigskip
The dynamical stability of Buchdahl's gaseous model may be assured by using the variational method [11,12] which states that the {\it sufficient} condition for the dynamical instability of a mass is that
the right-hand side of the following equation 

\begin{eqnarray}
\lefteqn{\sigma^2 \int_{0}^{R} e^{(3\lambda - \nu) /2} (P + E) r^2 \xi^2 dr = }  \nonumber  \\
  & & 4 \int_{0}^{R} e^{(\lambda + \nu)/2} r P'\xi^2 dr  \nonumber  \\
  & & + \int_{0}^{R} e^{(\lambda + 3 \nu)/2} [\gamma P/r^2] {(r^2 e^{-\nu/2} \xi)'}^2 dr  \nonumber  \\
  & & - \int_{0}^{R} e^{(\lambda + \nu)/2} [P'^2/(P + E)] r^2 \xi^2 dr  \nonumber  \\
  & & + 8\pi \, \int_{0}^{R} e^{(3\lambda + \nu) /2} P (P + E) r^2 \xi^2 dr
\end{eqnarray}

\bigskip

vanishes for some chosen `trial function' $\xi$ which
satisfies the boundary conditions
\begin{eqnarray}
\xi & = & 0 \hspace{.2in} {\rm at} \hspace{.2in} r = 0,
\end{eqnarray}
and
\begin{eqnarray}
\delta P & = & - \xi P' - \gamma P e^{\nu/2} [(r^2 e^{-\nu/2} \xi)'/r^2]  \nonumber  \\
         & = & 0 \hspace{.2in} {\rm at} \hspace{.2in} r = R
\end{eqnarray}

\bigskip

The system, consists of gaseous configurations discussed in section 2, is considered fully `isentropic' so that   the quantity $\gamma = [(P + E)/P](dP/dE) = \Gamma_1$ appearing in Eq.(13) represents the adiabatic index. The prime appearing in Eqs.(13) and (15)   denotes   radial 
derivative, $\delta P$ is called the `Lagrangian 
displacement  in  pressure', $\sigma$ is known as the angular frequency of pulsation and $ R $ is the size of the configuration. In order to solve Eq.(13) for the system under consideration, we employ the fourth-order Runge-Kutta iteration by using the `trial function' as a power series of the form $\xi = b_1 r(1 + a_1 r^2  + a_2 r^4  + a_3 r^6)e^{\nu/2}$, 
(where $a_1, a_2, \,{\rm and}\, a_3,$ are arbitrary constants)[10,13]. The positive values of the squared angular frequency of pulsation, ${\sigma_1}^2$, for all the permissible values of $u$  indicate the existence of pulsationally stable configurations in the range $0 < u \leq 0.20$ as shown in Table 1. For comparison, we repeat the calculations for the choice of the `trial function' $\xi = re^{\nu/4}$ used by Knutsen [6]. The positive values of the squared angular frequency of pulsation, ${\sigma_2}^2$, for all values of $u$ as shown in Table 1 confirm the above finding. We also calculate the fractional binding-energy $\alpha [\equiv (M_r - M)/M$; where $M_r$ is called the rest-mass of the entire configuration ([15])]. The positive values of $\alpha$ for all the above-mentioned values of $u$ show the existence of gravitationally bound structures.

\bigskip

\bigskip
\noindent {\large \bf 6.\,\,\, Results and Conclusions}
\bigskip

 The present study shows that the isentropic, perfect fluid subclass of Buchdahl's gaseous model is pulsationally stable and gravitationally bound for the range of compactness ratio in which the solution satisfies the condition of regularity. This finding might be helpful for the future investigation of the relativistic supermassive stars as models for quasars on the basis of this solution, together with the usual models of neutron stars.

\bigskip

{\large \bf \noindent Acknowledgments}

\bigskip

The author acknowledges Aryabhatta Research Institute of Observational Sciences (ARIES),
Nainital for providing library and computer-centre facilities.

\pagebreak
{\large \bf References}
\begin{itemize}

\item[]{1.}
     Kramer,  D. Stephani, H. Herlt E. and  MacCallum, M., 
     in: ``Exact Solutions of Einstein's Equations'',
     CUP, Cambridge (1980).
\item[]{2.}     
      Tolman,  R. C., {\it Phys. Rev.} {\bf 55}, 364 (1939). 
\item[]{3.} 
      Mehra, A. L., {\it J. Aust. Math. Soc. \/} {\bf 6}, 153 (1966).
\item[]{4.}
      Buchdahl, H. A., {\it Astrophys. J.} {\bf 147}, 310 (1967).
\item[]{5.}
      Schutz, B. F., A First Course in General Relativity (Cambridge University Press,      Cambridge) (1985).
\item[]{6.} 
      Knutsen, H., {\it Gen. Rel. Grav.} {\bf 20}, 317 (1988).
 \item[]{7.} Negi, P. S., \& Durgapal, M. C., {\it Astrophys. Space Sci.} {\bf 275}, 185 (2001). 
\item[]{8.} M. C. Durgapal and P. S. Rawat, {\it Mon. Not. R. Astr. Soc. \/} {\bf 192}, 659 (1980). 
\item[]{9.}
      Negi, P. S., \& Durgapal, M. C. 
      {\it Gen. Rel. Grav.} {\bf 31}, 13 (1999).
\item[]{10.}  
Tooper, R. F. , 
      {\it Astrophys. J.} {\bf 142}, 1541 (1965).
\item[]{11.} Chandrasekhar, S., {\it Phys. Rev. Letters} {\bf 12}, 114 \& 437 (1964).
\item[]{12.} Chandrasekhar, S. , {\it Astrophys. J.} {\bf 140}, 417 (1964).
\item[]{13.} Bardeen, J. M., Thorne, K. S., \& Meltzer, D. W., {\it Astrophys. J.} {\bf 145}, 505 (1966).
\item[]{14.} Caporaso, G., \& Brecher, K., {\it Phys. Rev. D} {\bf 20}. 1823 (1979).
\item[]{15.} Zeldovich, Ya. B., \& Novikov I. D., Relativistic  Astrophysics, Vol.1,
      Chicago Univ. Press, Chicago (1978).
\end{itemize}

\begin{table*}     
\caption{The angular frequency of pulsational, $\sigma_1$, for Buchdahl's gaseous models as calculated by the variational method (Eq.13) for the choice of the `trial function' $\xi = b_1 r(1 + a_1 r^2  + a_2 r^4  + a_3 r^6)e^{\nu/2}$, 
where $a_1 = (1/10R^2), a_2 = (1/5R^4), a_3 = (3/10R^6),$ are arbitrary constants, is shown in column 4 for different values of the compactness ratio $u$ (column 1) in the range $0 < u \leq 0.20$. For comparison, the corresponding angular frequency of pulsational, $\sigma_2$, calculated for the choice of the `trial function' $\xi = re^{\nu/4}$ in Eq.(13) is shown in column 5. The binding-energy per unit mass, $ \alpha[\equiv (M_r - M)/M, $ where $M_r$ is the rest mass of the configuration] is shown in column 3. Column 2 shows different values of the constant $k[=(1 - 2u)]$, used by Buchdahl[4] and Knutsen[6] for corresponding values of $u$.}
\begin{center}
\begin{tabular}{ccccc}

\hline
$u$ & $k$ & $\alpha$ & $(R\sigma_1)^2$ & $(R\sigma_2)^2$ \\

\hline

0.00005 & 0.99990  & 0.00002 & 0.00019 & 0.00019 \\
0.00050 & 0.99900 & 0.00025 & 0.00189 & 0.00191  \\
0.00500 & 0.99000 & 0.00251 & 0.01882 & 0.01910  \\
0.01000 & 0.98000 & 0.00505 & 0.03751 & 0.03812  \\
0.02000 & 0.96000 & 0.01021 & 0.07456 & 0.07601  \\
0.03000 & 0.94000 & 0.01547 & 0.11126 & 0.11382  \\
0.04000 & 0.92000 & 0.02085 & 0.14774 & 0.15172  \\
0.05000 & 0.90000 & 0.02635 & 0.18416 & 0.18993  \\
0.06000 & 0.88000 & 0.03198 & 0.22069 & 0.22872  \\
0.07000 & 0.86000 & 0.03775 & 0.25757 & 0.26840  \\
0.08000 & 0.84000 & 0.04365 & 0.29506 & 0.30936  \\
0.09000 & 0.82000 & 0.04971 & 0.33350 & 0.35211  \\
0.10000 & 0.80000 & 0.05592 & 0.37334 & 0.39730  \\
0.11000 & 0.78000 & 0.06230 & 0.41514 & 0.44578  \\
0.12000 & 0.76000 & 0.06887 & 0.45967 & 0.49873  \\
0.13000 & 0.74000 & 0.07562 & 0.50798 & 0.55778  \\
0.14000 & 0.72000 & 0.08260 & 0.56162 & 0.62536  \\
0.15000 & 0.70000 & 0.08977 & 0.62293 & 0.70518  \\
0.16000 & 0.68000 & 0.09718 & 0.69571 & 0.80346  \\
0.16667 & 0.66670 & 0.10227 & 0.75382 & 0.88444  \\
0.18000 & 0.64000 & 0.11280 & 0.91086 & 1.11389  \\
0.19000 & 0.62000 & 0.12105 & 1.10873 & 1.42120  \\
0.19995 & 0.60010 & 0.12958 & 1.81559 & 2.63263  \\

\hline
\end{tabular}
\end{center}
\end{table*}

\end{document}